\documentclass[aps,pre,twocolumn,superscriptaddress]{revtex4}

\usepackage{amsmath,amssymb,amsfonts,graphicx}
\usepackage{url}

\begin{document}

\title{Local box-counting dimensions of discrete quantum eigenvalue spectra: \\ Analytical connection to quantum spectral statistics}

\author{Jamal Sakhr}
\affiliation{Department of Physics and Astronomy, University of Western Ontario, London, Ontario N6A 3K7 Canada}
\author{John M. Nieminen}
\affiliation{Christie Digital Systems Canada Inc., 809 Wellington Street North, Kitchener, Ontario N2G 4Y7 Canada}

\date{\today}

\begin{abstract}
Two decades ago, Wang and Ong [Phys. Rev. A \textbf{55}, 1522 (1997)] hypothesized that the local box-counting dimension of a 
discrete quantum spectrum should depend exclusively on the nearest-neighbor spacing distribution (NNSD) of the spectrum. In this paper, we validate their hypothesis by deriving an explicit 
formula for the local box-counting dimension of a countably-infinite discrete quantum spectrum. This formula expresses the local box-counting dimension 
of a spectrum in terms of single and double integrals of the NNSD of the spectrum.
As applications, we derive an analytical formula for Poisson spectra and closed-form approximations to the local box-counting dimension for spectra having Gaussian orthogonal ensemble (GOE), Gaussian unitary ensemble (GUE), and Gaussian symplectic ensemble (GSE) spacing statistics. 
In the Poisson and GOE cases, we compare our theoretical formulas with the published numerical data of Wang and Ong  
and observe excellent agreement between their data and our theory. 
We also study numerically the local box-counting dimensions of the Riemann zeta function zeros and the alternate levels 
of GOE spectra, which are often used as numerical models of spectra possessing GUE and GSE spacing statistics, respectively. 
In each case, the corresponding theoretical formula is found to accurately describe the numerically-computed local box-counting dimension. 
\end{abstract}


\maketitle

\section{Introduction}

In fractal geometry \cite{BBM}, scale invariance is quantified through the 
so-called fractal dimension. A \emph{constant} fractal dimension is often defined as the scaling exponent 
of a geometric power law. The box-counting dimension is the simplest and most 
pervasive example of a fractal dimension (see, for example, Ref.~\cite{addison} for definitions and review). 
``Fractal analysis", which is a term that is frequently 
used in many different contexts throughout the sciences, in practice, often means to seek out scaling behavior 
and ultimately deduce a fractal dimension from a log-log plot of the box-counting function. 
Except for special mathematical sets such as the classical fractals, log-log plots of the box-counting function 
are never perfect straight lines. In other words, the 
box-counting dimension is generally non-constant and a function of the measurement or observation scale and is  
strictly constant only in special cases (e.g., the classical fractals) \cite{takabook}. 

The box-counting dimension has in fact quite often been observed to be a smooth or discontinuous function 
of the measurement scale. An example of the former case is fluid interfaces in turbulence 
\cite{CatBond00} and an example of the latter is fracture networks in geophysics \cite{discontbox}. 
There are however very few examples for which the 
scale dependence of the box-counting dimension is understood analytically. One well-known theoretical example is the family of statistical mechanical models involving randomly distributed spheres, rods, and disks \cite{Hamburger}. 


Interestingly, the box-counting dimensions of discrete quantum energy-level spectra have also been found to be smooth functions of the measurement 
scale \cite{WangOng}. The exact scale-dependent behaviors of the box-counting dimensions are not known analytically, 
and it is this particular problem that we wish to address in this paper. Before we begin, it is important to establish the background to this problem. 

In 1985, Cederbaum, Haller, and Pfeifer \cite{Cederbaum} (CHP) defined a scale-dependent generalization of the box-counting dimension, which they called the ``fractal dimension function'', that depends on both the measurement scale and the number of elements in a given set. They applied it to discrete quantum spectra and found that different spectra (of a prescribed length) 
had different scale-dependent behaviors and that certain statistical properties of spectra essentially determined the behavior of their scale-dependent fractal dimension. 
CHP were also able to derive an analytical formula for the ``fractal dimension function'' of a finite discrete quantum spectrum, which interestingly they found depended 
on the nearest-neighbor spacing distribution (NNSD) of the spectrum \emph{and} on the length of the spectrum 
(i.e., the number of energy levels in the sample). There are however serious problems that arise from the 
dependence on the number of energy levels and some of these problems were pinpointed and discussed twelve years later in a paper by Wang and Ong (WO) \cite{WangOng}.  

WO argued that, for a discrete eigenvalue spectrum, the function $D_b(r)$ defined in the next section [see Eq.~(\ref{effdim})], which is a formal scale-dependent generalization of the box-counting dimension, should only depend on the NNSD of the spectrum \footnote{Wang and Ong (borrowing from Takayasu \cite{takabook}) refer to the function $D_b(r)$ as the ``effective fractal dimension'', and in their paper (Ref.~\cite{WangOng}), it is denoted by $D_F(r)$.}. Although reasonable, their hypothesis lacked analytical proof in the sense that they could not give an explicit formula (exact or otherwise) for $D_b(r)$ in terms of the spacing distribution.  
WO computed $D_b(r)$ numerically for spectra having Poisson and Wigner 
spacing distributions and also for the vibrational spectra of the $\text{SO}_2$ molecule. 
The authors also stated that box-counting methods are amenable to
numerical implementation, but ``will encounter difficulties 
when one attempts to search for an analytical solution''. There is, in principle, no difficulty in seeking out an 
analytical solution, and in this paper, we will derive an exact general formula for the (scale-dependent) 
box-counting dimension $D_b(r)$ of a discrete quantum spectrum using only 
elementary results from the statistical theory of spectra \cite{Mehta}. 
As we shall see, $D_b(r)$ does indeed depend only on the NNSD of the spectrum. 

\section{Local Box-Counting Dimension}\label{definnys}

To define the box-counting dimension, we first consider a uniform partition 
of the embedding space of a given set into a grid of non-overlapping
boxes of side-length $r$. (The embedding space for an energy spectrum
is $\mathbb{R}$, in which case the boxes are actually line segments
of length $r$.) This grid of boxes is often called an $r$-mesh. Let
$N(r)$ denote the number of boxes needed to cover the set, that is, the
number of boxes that have a non-empty intersection with the set. The box-counting dimension $D_b$ of a set can be defined as follows \cite{addison,takabook}: 
\begin{equation}\label{bcopfordefn}
D_b=-\lim_{r\to0}{d\log_a[N(r)]\over d\log_a(r)}, 
\end{equation}
where the base $a>1$ is arbitrary. For a given set, the box-counting function $N(r)$ may or may not follow a power law of the form 
\begin{equation}
N(r)\sim r^{-D_b}. 
\end{equation} 
In the latter case, the given set possesses a scale-dependent geometry and the interest then lies in understanding the behavior of the  
box-counting dimension as a function of the measurement or observation scale $r$. In other words, the object of interest in such cases is the function
\begin{equation}\label{effdim}
D_b(r)=-{d\log_a[N(r)]\over d\log_a(r)}.
\end{equation}
The above function is well-defined provided $N(r)$ is a smooth function of the measurement scale. 
Note that $D_b(r)$ will be constant only when $N(r)$ is an exact power law and is non-constant otherwise. 
The function $D_b(r)$ has several different names in the physics literature. For instance, in the well-known and
often cited text by Takayasu \cite{takabook}, it is referred to as the
``effective fractal dimension'', so-called after Mandelbrot, who in his
book \cite{BBM} discussed the notion of an ``effective dimension'' 
that depends on the resolution or the scale of measurement. 
We shall refer to $D_b(r)$ as the local box-counting dimension, 
which is the terminology used in Ref.~\cite{Bologna}. 

\section{Discrete Eigenvalue Spectra} 

Suppose we are given a countably-infinite
discrete spectrum for which the spacing $s$ between adjacent energy
levels is described by a probability density function $P(s)$ with mean $\bar{s}=\int_0^\infty sP(s)ds$. Consider
the subset of the spectrum that lies in the interval $\left[\mathsf{E}_{\text{min}},\mathsf{E}_{\text{max}}\right]$ and
partition this interval into $(\mathsf{E}_{\text{max}}-\mathsf{E}_{\text{min}})/r \equiv L/r$ intervals (boxes) of size $r$. Let
$N(r)$ denote the number of these intervals that contains one or more
eigenvalues. The fraction $N(r)/(L/r)=(r/L)N(r)$ of boxes that contain
eigenvalues is equivalent to the probability $Q(r)$ that one of
the boxes chosen at random contains one or more eigenvalues. This
probability can alternatively be expressed as: 
$Q(r)=1 - E(r)$, where $E(r)$ is the
probability that an arbitrary interval of length $r$ 
(i.e., a box of length $r$ chosen at random) contains {\em no}
eigenvalues. Thus, the
number of boxes needed to cover the subset of the spectrum lying in $\left[\mathsf{E}_{\text{min}},\mathsf{E}_{\text{max}}\right]$ 
is the number of intervals $L/r$ multiplied by $Q(r)$:
\begin{equation}\label{Eqn:NumbBox}
    N(r) = \frac{L}{r}[1 - E(r)].
\end{equation}
Using definition (\ref{effdim}) and the fact that 
\begin{equation} \label{Eqn:boxdim1} 
    \log_a[N(r)] = \log_a(L) - \log_a(r) + \log_a[1 - E(r)],
\end{equation}
the local box-counting dimension $D_b(r)$ for a discrete energy-level
spectrum is therefore  
\begin{equation} \label{Eqn:boxdim2}
    D_b(r) =1 + \frac{r}{[1 - E(r)]} \frac{dE(r)}{dr}.
\end{equation}
All that remains to determine is $E(r)$ and $dE(r)/dr$.
Given $P(s)$, it is a relatively simple matter to write down an expression 
for $E(r)$. 

In the statistical theory of spectra, the function $E(x)$ is known as the ``gap probability'' 
\footnote{In the notation of Ref.~\cite{Mehta}, the function $E(n;x)$ denotes the
probability that an arbitrary interval of length $x$ contains $n$ eigenvalues. 
In this paper, we use $E(x) \equiv E(0;x)$.}. The link between $E(x)$ and $P(s)$ can 
be quickly established using elementary results from the statistical theory of spectra. 
We shall here simply quote the following identity from Ref.~\cite{Mehta}: $dE(x)/dx=-(1/\bar{s})F(x)$,
where $F(x)$ is the probability that there are no eigenvalues within a distance $x$ of 
an eigenvalue chosen at random, or equivalently, the probability that the distance to the
nearest neighbor is \emph{greater} than $x$ \footnote{In the notation of Ref.~\cite{Mehta}, this identity is written as: $-dE(0;x)/dx=F(0;x)$, where $F(n;x)$ is the probability that there are $n$ eigenvalues within a distance $x$ of some eigenvalue chosen at random. In this paper, we use $F(x) \equiv F(0;x)$. Note that the derivation of the above-quoted identity from Ref.~\cite{Mehta} implicitly assumes a unit mean spacing (i.e., $\bar{s}=1$). We make no such assumption here, and it can be shown that removing this assumption introduces the additional given factor of $\bar{s}$.}. The complementary probability $\Psi(x) \equiv [1-F(x)]$ 
is the probability that the nearest neighbor (to an eigenvalue chosen at random) 
is within a distance $x$. In other words, $\Psi(x)$ is the probability that the nearest-neighbor 
distance is \emph{less} than or equal to $x$, which, by definition, is given by $\int_0^x P(s)ds$.
Combining these two results yields the following identity relating $E(x)$ and $P(s)$:
\begin{equation} \label{Eqn:FTC}
    {dE(x)\over dx}=-\left({1\over\bar{s}}\right)[1-\Psi(x)]=-\left({1\over\bar{s}}\right)\int_x^\infty P(s)ds. 
\end{equation}
Integrating Eq.~(\ref{Eqn:FTC}), and noting that $E(0)=1$, 
the probability that an arbitrary interval of length $r$ does not
contain any eigenvalues is 
\begin{equation} \label{Eqn:E_gen}
    E(r) = 1 - \left({1\over\bar{s}}\right)\int_0^{r} \left[ 1-\Psi(x) \right] dx.
\end{equation}
Substituting (\ref{Eqn:FTC}) and (\ref{Eqn:E_gen}) into (\ref{Eqn:boxdim2}) then immediately yields the following
general formula for $D_b(r)$ in terms of $P(s)$:
\begin{equation} \label{Eqn:boxdim3}
    D_b(r)=1-\frac{r\int_r^\infty P(s)ds}{\int_0^r\int_x^\infty P(s)ds dx}.
\end{equation}
The above formula is the main theoretical result of this paper 
\footnote{Interestingly, a formula analogous to Eq.~(\ref{Eqn:boxdim3}) that connects scale distributions 
and coverage dimensions of level sets arising in turbulence has been obtained by Catrakis and Dimotakis \cite{CatPRL96}.}. 
Note that although $N(r)$ depends on $L$ (and hence on the
size of the sample spectrum), $D_b(r)$ is independent of $L$. 
Armed with formula (\ref{Eqn:boxdim3}), we are now in a position to derive the local box-counting dimension of specific sequences of levels. 

\section{Applications}

\subsection{Poisson Spectra}

\begin{figure} 
\scalebox{0.6}{\includegraphics*{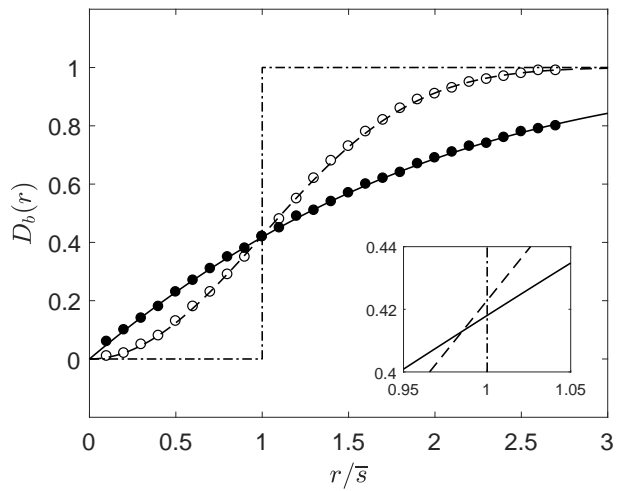}}
\caption{\label{newGOE} The local box-counting dimension $D_b(r)$ 
versus $r/\bar{s}$ for energy-level spectra having Poisson and Wigner spacing distributions. 
The solid and open circles are the numerical values of the ``effective fractal dimension'' obtained by WO \cite{WangOng} for levels having Poisson and Wigner spacing distributions, respectively. 
The solid and dashed curves are the theoretical $D_b(r)$ formulas (\ref{Eqn:P_box}) and (\ref{Eqn:W1_box}) obtained for levels having Poisson and Wigner spacing distributions, respectively. 
The dashed-dotted lines are for reference only and correspond to the special case of equally-spaced levels.
The inset shows a close-up view around the intersection of the two theoretical curves from which it is obvious that the
Poisson and Wigner curves do not intersect at $r=\bar{s}$ (contrary to 
the observations of WO \cite{WangOng}).}
\end{figure}

The NNSD for an ordered sequence of independent random levels is 
given by \cite{Mehta}
\begin{equation} \label{Eqn:P_P}
    P_P(s) = (1/\bar{s})\exp(-s/\bar{s}), 
\end{equation}
where $\bar{s}$ is the mean spacing. Substituting (\ref{Eqn:P_P}) into (\ref{Eqn:boxdim3}) and performing the straightforward algebra yields 
\begin{equation} \label{Eqn:P_box}
    D_{b}(r) = 1 - {(r/\bar{s})\exp(-r/\bar{s})\over{1-\exp(-r/\bar{s})}}.
\end{equation}
This is displayed in Fig.~\ref{newGOE}, and 
for comparison, we also plot the ``effective fractal dimension''
data obtained numerically by WO for a Poisson spectrum.
Note that 100,000 levels were used by WO to numerically compute $D_b(r)$. The WO data is in excellent agreement with formula (\ref{Eqn:P_box}).

\subsection{GOE Spectra}

We seek here to obtain a closed-form approximation  
to $D_b(r)$ in the case of spectra that follow Gaussian orthogonal ensemble (GOE) statistics. In order to do so, we shall use the Wigner surmise for the GOE: 
\begin{equation} \label{Eqn:W1_P(S)}
     P_W(s; \beta=1) = {\pi\over2\bar{s}}\left({s/\bar{s}}\right)\exp\left(-{\pi\over4}\left({s/\bar{s}}\right)^2\right).
\end{equation}
Although the above distribution (commonly referred to as the Wigner distribution) is an exact result only for real symmetric $2 \times 2$ random 
matrices, it 
serves as an excellent analytical approximation to the asymptotic 
Mehta-Gaudin distribution appropriate for artbitrarily large random
matrices from the GOE \cite{Mehta}.
Substituting (\ref{Eqn:W1_P(S)}) into (\ref{Eqn:boxdim3}) and performing the necessary algebra yields 
\begin{equation} \label{Eqn:W1_box}
    D_{b}(r) = 1 - {(r/\bar{s})\exp\left(-{\pi\over4}(r/\bar{s})^2\right)\over\text{erf}\left({\sqrt{\pi}\over2}(r/\bar{s})\right)}, 
\end{equation}
where $\text{erf}(z)$ is the error function \cite{AS}. 
This result is displayed in Fig.~\ref{newGOE} along with the ``effective
fractal dimension'' that was calculated numerically by WO~\cite{WangOng}
for a spectrum possessing a Wigner spacing distribution (\ref{Eqn:W1_P(S)}). Once again, the WO data is in excellent agreement with formula (\ref{Eqn:W1_box}).

\subsection{GUE Spectra}

In the case of spectra that follow Gaussian unitary ensemble (GUE) statistics, a non-elementary closed-form 
approximation to $D_b(r)$ can be obtained by using the Wigner surmise for the GUE:
\begin{equation} \label{Eqn:W2_P(S)}
     P_W(s;\beta=2)={32\over\pi^2\bar{s}}(s/\bar{s})^2\exp\left(-{4\over\pi}(s/\bar{s})^2\right).
\end{equation}
Although (\ref{Eqn:W2_P(S)}) is exact only for Hermitian $2\times2$ random 
matrices, it serves as an excellent analytical approximation to the asymptotic 
Mehta-Gaudin distribution appropriate for artbitrarily large random
matrices from the GUE \cite{Mehta}.
Substituting (\ref{Eqn:W2_P(S)}) into (\ref{Eqn:boxdim3}) and performing the necessary algebra yields
\begin{widetext}
\begin{equation} \label{Eqn:W2_box}
D_{b}(r) = 1 - {(r/\bar{s})\bigg(\text{erfc}\left({2\over\sqrt{\pi}}(r/\bar{s})\right)+{4\over\pi}(r/\bar{s})\exp\Big(-{4\over\pi}(r/\bar{s})^2\Big)\bigg)\over1-\exp\Big(-{4\over\pi}(r/\bar{s})^2\Big)+(r/\bar{s})\text{erfc}~\left({2\over\sqrt{\pi}}(r/\bar{s})\right)}, 
\end{equation}
\end{widetext}
where $\text{erfc}(z)$ is the complementary error function \cite{AS}. 

\begin{figure} 
\scalebox{0.6}{\includegraphics*{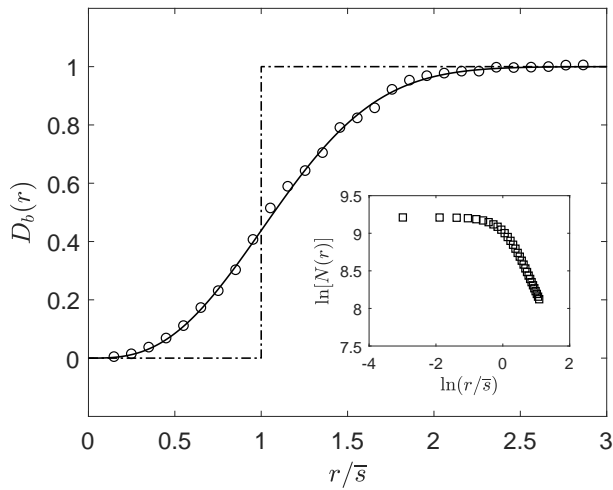}}
\caption{\label{GUE_Riemann} $D_b(r)$ versus $r/\bar{s}$ for 10,000
high-lying zeros of the Riemann zeta function. The open circles
are numerical derivatives calculated from the $\ln[N(r)]$ versus $\ln(r/\bar{s})$ 
data, which is shown as open squares in the inset. 
The solid curve is the theoretical $D_b(r)$ formula (\ref{Eqn:W2_box}) obtained for levels having GUE spacing statistics. 
The dashed-dotted lines are for reference only and correspond to the special case of equally-spaced levels.}
\end{figure}

As a numerical example, we examine the local box-counting dimension 
of 10,000 high-lying zeros of the Riemann zeta function, in particular, 
the ($10^{22}+1$)th zero to the ($10^{22}+10^4$)th zero \cite{OdlyzkoTable}. 
It is conjectured that (asymptotically) the zeta zeros have the same statistical properties as 
the eigenvalues of arbitrarily large random matrices from the GUE and numerical observations so far corroborate this conjecture \cite{Odlyzko}. We computed the local box-counting dimension of the above-specified $10,000$ zeros 
by calculating the average slope at each point of the numerically-determined 
$\ln[N(r)]$ versus $\ln(r/\bar{s})$ plot. This is shown in Fig.~\ref{GUE_Riemann}. 
The theoretical $D_b(r)$ curve [Eq.~(\ref{Eqn:W2_box})] 
accurately describes the numerically-computed local box-counting dimension.

\subsection{GSE spectra}

In the case of spectra that follow Gaussian symplectic ensemble (GSE) statistics, a non-elementary closed-form 
approximation to $D_b(r)$ can be obtained by using the Wigner surmise for the GSE:
\begin{equation} \label{Eqn:W4_P(S)}
     P_W(s;\beta=4)={2^{18}\over3^6\pi^3\bar{s}}(s/\bar{s})^4\exp\left(-{64\over9\pi}(s/\bar{s})^2\right). 
\end{equation}
Substituting (\ref{Eqn:W4_P(S)}) into (\ref{Eqn:boxdim3}) and performing the necessary algebra yields 
\begin{widetext}
\begin{equation} \label{Eqn:W4_box}
    \displaystyle D_{b}(r) = 1 - {(r/\bar{s})\bigg(\text{erfc}\left({8\over3\sqrt{\pi}}(r/\bar{s})\right)+\Big({16\over3\pi}(r/\bar{s})+{2048\over81\pi^2}(r/\bar{s})^3\Big)\exp\Big(-{64\over9\pi}(r/\bar{s})^2\Big)\bigg)\over1-\Big(1+{16\over9\pi}(r/\bar{s})^2\Big)\exp\Big(-{64\over9\pi}(r/\bar{s})^2\Big)+(r/\bar{s})\text{erfc}\left({8\over3\sqrt{\pi}}(r/\bar{s})\right)}. 
\end{equation}
\end{widetext} 

\begin{figure} 
\scalebox{0.6}{\includegraphics*{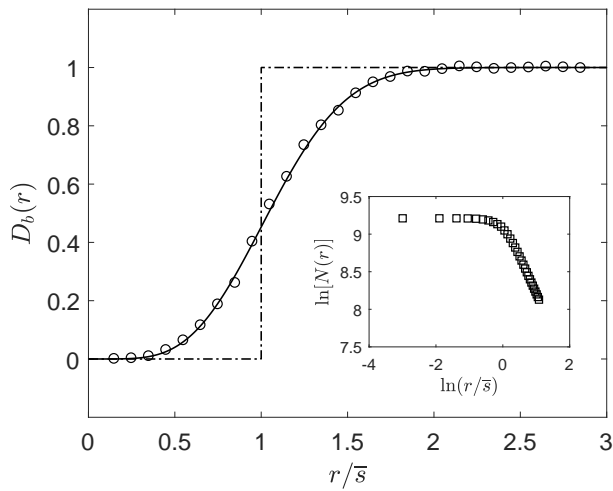}}
\caption{\label{GSE_fig} $D_b(r)$ versus $r/\bar{s}$ for 10,000
alternate levels of a GOE spectrum (of length 20,000). The open circles
are numerical derivatives calculated from the $\ln[N(r)]$ versus $\ln(r/\bar{s})$ 
data, which is shown as open squares in the inset. 
The solid curve is the theoretical $D_b(r)$ formula (\ref{Eqn:W4_box}) obtained for levels having GSE spacing statistics. 
The dashed-dotted lines are for reference only and correspond to the special case of equally-spaced levels.}
\end{figure}

As a numerical example, we computed the local box-counting dimension 
of $10,000$ alternate levels of a GOE spectrum (consisting of 20,000 eigenvalues) by calculating the 
average slope at each point of
the numerically-determined $\ln[N(r)]$ versus $\ln(r/\bar{s})$ plot. This is shown in Fig.~\ref{GSE_fig}. 
As before, the theoretical $D_b(r)$ curve [Eq.~(\ref{Eqn:W4_box})] 
accurately describes the numerically-computed local box-counting dimension.

\section{Conclusion}

To summarize, we have provided an analytical theory for the local 
box-counting dimension of discrete quantum eigenvalue spectra. 
Our main formula [Eq.~(\ref{Eqn:boxdim3})] explicitly shows that the 
local box-counting dimension of a discrete spectrum depends only on its NNSD, as was
first hypothesized two decades ago by Wang and Ong \cite{WangOng}. 
In fact, according to Eq.~(\ref{Eqn:boxdim3}), the local box-counting dimension of a 
discrete spectrum is simply an integral transformation of its NNSD.

As applications of our theory, we derived an analytical formula for Poisson spectra and closed-form 
approximations to the local box-counting dimension for spectra having GOE, GUE, and GSE spacing statistics.
In the Poisson and GOE cases, we compared our theoretical formulas with the published numerical data of WO \cite{WangOng} 
and observed impeccable agreement between their data and our theory. 
We also studied numerically the local box-counting dimensions of the Riemann zeta function zeros and the alternate levels 
of GOE spectra, which are often used as numerical models of spectra possessing GUE and GSE spacing statistics, respectively. 
In each case, the corresponding theoretical formula [Eq.~(\ref{Eqn:W2_box}) for GUE spectra and Eq.~(\ref{Eqn:W4_box}) for GSE spectra] 
was found to accurately describe the numerically-computed local box-counting dimension. 

Quantum eigenvalue spectra might appear to be just another mathematical playground for the 
tools of fractal geometry, and from the standpoint of the more general discussion given in the Introduction, discrete quantum 
spectra are indeed cited as just another example of point sets having a scale-dependent geometry. However, 
in quantum mechanics itself, the geometric scaling properties of energy-level spectra have a much more profound significance 
(see, for example, Refs.~\cite{PRL97,MPL99andRMP99,thermowithstats,coron12,fibo12}). 
So, unlike in many lines of research where fractal geometry has been adopted mainly as a descriptive tool, applying the concepts of fractal 
geometry to quantum eigenvalue spectra are not merely for descriptive purposes. 

We conclude with the following clarification, which is important in the above context.  
WO think of discrete spectra as ``sets of a series of discrete points that exhibit fractal properties'', 
or more simply as, ``fractal sets''. Unless it is made absolutely clear \emph{in what sense} discrete spectra are ``fractal sets'', it is ambiguous (and even wrong) to refer to them as such \footnote{For an example of how this type of question can be answered quite rigorously, 
see Ref.~\cite{hydrogenspect}, which gives one unambiguous answer to the question: in what sense are 
the spectral lines of hydrogen (which are not all discrete) a fractal set.}.  
By merely computing $D_b(r)$ for a discrete spectrum, one has not (to quote WO) 
``investigated the fractal properties'' of the spectrum. 
In fact, the numerical studies of WO clearly demonstrate that discrete quantum spectra do not have box-counting type scaling behavior (i.e., no ``fractal structure" 
defined in terms of a constant box-counting dimension) and thus have a scale-dependent geometry. 

\begin{acknowledgments}
We thank A.M. Odlyzko for open access to the zeta function zeros. 
\end{acknowledgments}

\end{document}